\documentstyle[11pt]{article}

\begin{document}

\begin{flushright}
Liverpool Preprint: LTH 349\\
 hep-lat/9506009\\
 5 June 1995\\
 \end{flushright}

\vspace{5mm}
\begin{center}
{\LARGE\bf
Is the Quenched Spectrum in agreement with Experiment?
}\\[10mm] {\large\it UKQCD Collaboration}\\[3mm]

{\bf  P. Lacock and  C. Michael}\\

{DAMTP, University of Liverpool, Liverpool, L69 3BX, U.K.}\\

\end{center}

\begin{abstract}

We analyse the meson  spectrum in quenched QCD  using lattice gauge
theory. By studying hadron propagation  with a variety of  operators
(both smeared and local),  we are able to extract the ground state and
first excited state masses with confidence.  We pay attention to the
correlations among the data used in  the fits to extract these masses
and couplings. We compare the  resulting hadron spectrum with experiment
and find evidence for  a significant departure in the pseudoscalar and
vector meson  masses.

\end{abstract}

\section{Introduction}

Lattice gauge theory is able to evaluate the hadron spectrum from  first
principles. In practice, the quenched approximation is needed  to keep
the computational resource manageable. Thus it is  valuable to calibrate
the quenched  approximation. Since data from different fermionic
actions and different lattice spacings need to be combined, it is
worthwhile to look for observables which are independent. This is
the motivation for the Edinburgh plot of $m_N/m_{\rho}$ versus
$m_{\pi}/m_{\rho}$ for example. Results for the baryon spectrum
are less precise than those for mesons, partly because the baryonic
states are heavier. Thus it will be helpful to construct combinations
of mesonic masses (pseudoscalar P and vector V in particular) to compare
with experiment. Typically, $m_{\rho}$ is used to set
the lattice spacing scale $a$ and $m_V/m_P$ is used to fix the quark
mass. The only remaining information is on the relative
quark mass dependence of the vector and pseudoscalar mesons. We thus
focus on this and propose to use
\begin{equation}
    J = m_{K^*} {d m_V \over d m_P^2}
\end{equation}
 as a measure. This is in the spirit of the Maiani-Martinelli
proposal~\cite{mm, ukqcd} to use $(m_{K^*}-m_{\rho})/(m_K^2 -
m_{\pi}^2)$ as an independent way to  determine the lattice scale.  We
also follow the suggestion of the GF11 group~\cite{GF11} to  avoid the
errors inherent in a chiral extrapolation by making the comparison at
the  quark  mass which reproduces  the mass of the K and K* mesons. As
well as the vector and scalar mesons, results  for the $J^{PC}=1^{++}$
axial mesons are also compared to experiment.

 We present our approach in the context of the quark model in section 2.
The new lattice results for the  meson masses  are then discussed in
section 3. After a comparison with other lattice  determinations in
section 4, we present our conclusions.

\section{The quark model}

    We shall study hadronic states using lattice techniques for  a range
of quark masses. Thus we need to establish mass formulae  for the
dependence of the mass of a meson on the masses ($q_1$ and $q_2$) of
the quarks  of which it is composed.

    The simplest assumption is that the meson masses are given by a
Taylor series in these quark masses. Assuming that the linear
approximation  is sufficient, for the vector mesons, this implies  that
\begin{equation}
  m_V(q_1,q_2) = d + c (q_1+q_2)
\end{equation}
Here the coefficient of $q_1$ and of $q_2$ is the same from symmetry.

   For the pseudoscalar mesons, chiral symmetry considerations
imply that the appropriate expression is
\begin{equation}
 m_P^2(q_1,q_2) = b (q_1+q_2).
\end{equation}

We shall be concerned here with light quarks: u, d and s. To  avoid
consideration of the small isospin violating effects, we  use the
isospin averaged mass values and treat the u and d  quarks as having the
same mass (non-strange: ns). Then to  have a check of these mass
formulae, one needs to consider the  mixing of the experimental states
with isospin 0. Taking the  usual assumption that the $\phi$ meson is
purely composed of s quarks while the $\omega$ meson has purely
non-strange quarks (and so is  degenerate in mass with the $\rho$
meson),  these relations yield an  equal mass splitting between $\rho$,
K$^*$  and $\phi$ mesons. This is in close agreement with data since the
mass differences of $\phi - {\rm K}^*$ and ${\rm K}^* -\rho$ are  equal
within 4\%. Note, however, that the differences ${\rm K}^* -\rho$ and
${\rm K}^* -\omega$ differ by 10\% which shows that the assumption  of
ideal mixing  of the $\phi$ and $\omega$ is not perfect.

Further checks of these mass formulae can be made from lattice  studies
where the quark masses can be adjusted freely. The  most accurate
lattice spectroscopy is currently available in the  quenched
approximation. Although the chiral  behaviour in the quenched
approximation will be modified~\cite{sharpe,  ber-golt}, this effect is
expected to be significant only at very  small quark masses. Thus
it is appropriate to compare the above mass formulae with quenched
lattice mass values without extrapolating to the chiral limit. In order
to set a common reference quark mass we shall choose
$$
   m_V = 1.8 m_P
$$
which corresponds to the experimental ${\rm K}^* / {\rm K}$ mass ratio.
We define the values of the vector and pseudoscalar masses at this
quark mass as $m_v$ and $m_p$ respectively.

Equations 1 and 2 imply that a plot of $m_V$ against $m_P^2$ will  be
a straight line for varying quark masses. This has been checked in
lattice studies where  a range of values of $q_1$ and $q_2$ are used.
Moreover, studies  with $q_1 \ne q_2$ have also been
conducted~\cite{ukqcd,LANL} and  they show excellent agreement with the
mass formulae above. A compilation of data at $\beta=6.0$ for Wilson
fermions is shown in fig.~1 and at $\beta=6.2$ for the clover fermionic
action in fig.~2.

In lattice studies, it is usual to relate the quark mass to bare
lattice parameters (eg. it is assumed to be linear in $1/K$ for  Wilson
fermions where $K$ is the Wilson hopping parameter).  Furthermore the
scale of masses involves the  lattice spacing $a$. Thus the pseudoscalar
and vector meson masses  are commonly used to fix the relationship of
$K$ to the quark  mass  and to fix $a$.  This fixing procedure tends to
obscure the  underlying agreement between the experimental spectrum and
the lattice  spectrum.  Here we wish to emphasize that there is one
combination  of the parameters in the above mass formulae which is
independent of  $a$ and of $K$ and which does not involve extrapolation
to the  chiral limit.  This is the dimensionless combination
\begin{equation}
    J=m_v {c \over b} = m_v {d m_V \over d m_P^2}
\end{equation}

Thus, as was emphasized by the GF11 group~\cite{GF11},  one can compare
lattice results with the experimental spectrum  without having to
extrapolate to the chiral limit.  The  combination $J$ can be determined
in principle by lattice results at two  quark masses. Thus one
parametrises the variation  of $m_V$ with $m_P^2$ as a straight line.
Then the intersection of this line with $m_V=1.8 m_P$ determines the
reference  value $m_v$ which is to be multiplied by the  slope to  yield
$J$.  The useful feature of  $J$ is that it is independent of the quark
mass values chosen to  determine it, provided that the relationship
of $m_V$ to $m_P^2$ is linear for those quark masses.

 We can use the  experimental spectrum of $\pi$, K, $\rho$, $\omega$,
K$^*$ and  $\phi$ mesons with masses 138.0, 495.7, 770, 781.9, 892.1,
1019.5 MeV respectively to  determine the coefficients in the mass
formulae.  These values are  consistent with the mass formulae, treating
the mesons as made  from  two types of quark: strange $q_s$ and
non-strange $q_{ns}$. One  finds $q_{ns}/q_s =0.04$. We use the K$^*$,
$\rho$ and K, $\pi$ differences to determine  $dm_V/dm_P^2$ since this
does not  depend on mixing with the isoscalar states. This  gives the
dimensionless combination
 $$
J=0.48(2).
$$
The error on this has been estimated from the $\phi - {\rm K}^*$ to
${\rm K}^* -\rho$ mass ratio which is 1.04\ .

We shall also study the $1^{++}$ mesons. The observed states  are the
a$_1$ meson of mass  1230(40) MeV and the f$_1$ mesons of mass 1285(5)
and 1427(2) MeV. If we assume that the heavier f$_1$ meson is
predominantly $s\bar{s}$,  then the strange partner K$^{*}_A$ would have
mass 1328(20) MeV. The experimental  situation regarding the strange
1$^+$ mesons is confused by the mixing  between the mesons from the
1$^{++}$ and 1$^{+-}$ nonets to give  two observed states with masses
1273(7) and 1402(7) MeV. In the absence of a predictive model for this
mixing, we take the above value of 1328(20) MeV as the mass of the
K$^{*}_A$.  Thus at $m_V=1.8 m_P$, experiment gives the ratio
$m({\rm K}^*_A)/m({\rm K}^*_V)= 1.48(2)$.

We now discuss quenched lattice determinations of the meson spectrum.

\section{Lattice Measurements}

The results we shall present come from a study of  light quark
propagators in the 60 configurations of size $24^3 \times 48$  at
$\beta=6.2$ obtained by UKQCD~\cite{ukqcd}. These are at hopping
parameters  $K=0.14144$  and $K=0.14226$ using the clover improved
fermion action. The new feature of the present analysis is that the
quark propagators  were determined with smeared sources and sinks.  The
Jacobi smearing  algorithm was used~\cite{smear} with $K_S=0.190$  and
$N=90$.  This is a gauge invariant smearing  prescription. The notation
we shall use is  that SL corresponds to (all) quarks smeared at the
source but  local at the sink, etc.  We measured SS and SL correlations
and  previous results~\cite{ukqcd} for the LL correlations were also
used in the  analysis.

For the mesonic operators at source and sink, we used several  spin
combinations. For the pseudoscalar mesons, both the  usual
pseudoscalar operator (P) $\overline{q} \gamma_5 q$  and the axial
operator (A) $\overline{q} \gamma_5\gamma_4 q$ were used.
Moreover we employed both P and A operators at both source and
sink, so yielding 4 combinations PP, AA, PA and AP. Since this was
carried out for the three smearing choices, this yielded 12
different quantities which could be used to explore the pseudoscalar
meson spectrum and couplings. For the vector meson channel we found
that extra spin combinations (i.e. using $\gamma_i \gamma_4$ as
well as the usual $\gamma_i$)  did not help since they had very
large statistical errors. For the axial vector meson channel, we
used the axial current as an operator and analysed results
with only a smeared source but local and smeared sink.

The advantage of having many operators available is that it
makes the fit to extract the mass values much more tightly
constrained. This can be seen since
the euclidean time formulation implies that eigenstates of the
transfer matrix contribute to hadronic correlators
$<\!H^a(0)H^b(t)\!>$ as
\begin{eqnarray}
 C_{ab}(t) & = & <\!H^a(0)H^b(t)\!>  \nonumber \\
& = &  h_0^a h_0^b (e^{-m_0t}+e^{-m_0(L-t)}) +
   h_1^a h_1^b (e^{-m_1t}+e^{-m_1(L-t)}) + \dots
 \end{eqnarray}
 \noindent where $h_i^c$ is  the amplitude to produce eigenstate $i$
from operator $H^c$.  The expression is written assuming periodic
boundary conditions in time for bosonic eigenstates - the AP  and PA
spin combinations for pseudoscalar mesons will need anti-periodic time
boundary conditions.

Because of the factorisation of the coefficients above, if $D$
different operators are used at source and sink (so that
$a,b=1,\ldots , D$ above) and $T$ different $t$-values are fitted, there
will be  ${1 \over 2}D(D+1)T$ observables to fit with $(D+1)M$
parameters, where $i=1,\ldots ,M$ eigenvalues are retained. Thus the
number of observables increases faster than the number  of parameters as
$D$ increases.  This makes it feasible to retain more eigenvalues,
which, in turn, allows a larger $t$-region to be fitted. Furthermore,
the fitted $t$-region extends to smaller $t$-values where the relative
errors are smaller.

 This conclusion that several observables should be studied
simultaneously  can also be motivated by noting that the ground state
is only  determined accurately when an estimate of the first excited
state  is available. This is necessary since the energy difference
controls the rate  of approach of $C_{ab}(t)$ to the expression given
by the ground state   component alone. However, fitting 2 (or more)
exponentials to just  one function  $C(t)$ is not very stable: better
is to have several such functions (provided that they do indeed have
different  relative amounts of ground state and excited state).

The advantage of measuring several independent observables has long
been known in the  lattice gauge theory community. It is the root of
the success of  the variational method and is {\it de rigeur} for
studies of  glueballs and potentials. The variational method has also
been used to  study the light quark spectrum~\cite{smear}. Because  the
data for different observables and for different  $t$-separations are
very highly correlated statistically, we prefer  to use a general
correlated fitting program in the present work. The advantages of
improved  determination of the spectrum are still retained.

The correlations among the hadronic Green functions $C_{ab}(t)$ are
significant, particularly for adjacent values of $t$. As discussed
in ref~\cite{cmam}, this has important consequences for estimating
the goodness of fit. Since we fit, for the pseudoscalar channel,
over 200 data points simultaneously, we cannot use a full
correlated $\chi^2$ approach since there are only 60 independent
configurations available. Methods for accommodating the essence of
the correlation among the data, while using less parameters, have
been proposed~\cite{cmam}. Here we use the method of the 5-diagonal
approximation to the inverse correlation matrix as our basic tool.
This method~\cite{cmam} copes with the strong correlations in $t$ very
economically - it uses only two	 parameters for each type of
observable.  To check on possible systematic errors, we use two
more extreme assumptions as a guide. These are a completely
uncorrelated fit and a fit allowing correlations among the data
in both $t$ and between different observables (ie $a$ and $b$ in
our notation). This latter approach used a smoothing of the
full correlation matrix by retaining the 12 largest eigenmodes and
replacing the eigenvalues of the remaining modes by their average value
- see ref~\cite{cmam}.

Errors were  obtained by taking bootstrap samples of configurations from
the  original 60. For quantities involving fits from different  quark
masses ($K$-values) or for different quantum numbers (P, V or A),  we
used the same set of bootstrap configurations so that a  bootstrap error
analysis of the final quantity of interest was  obtained.

For each meson, we used a two-state fit. As discussed above, one state
fits only are acceptable for a limited range of $t$ (i.e. larger $t$),
while we found three state fits to be unstable. The $t$-range was chosen
by requiring the $\chi^2$ per degree of freedom of our basic fit  to be
acceptable (i.e. close to 1.0). Because the evaluation of the  goodness
of fit can be difficult with highly correlated data, we are  aware that
the appropriate $t$-range might be somewhat wider or  narrower. Since
including smaller $t$-values tends to increase the  fitted mass values,
we took account of this source of systematic errors  by varying the
$t$-range fitted.

The results for the two state fits to the pseudoscalar and vector
mesons are presented in Table~1 and shown in figures 3 to 7. These
results are compatible within  the quoted errors with the earlier
analysis~\cite{ukqcd} using  local operators and making one state fits
(i.e. plateau fits in  the effective mass) to much smaller $t$-ranges.
For the axial mesons, we found that a two state fit was not very stable,
so we fixed the mass of the excited state at 1 GeV. The results are
shown in Table~1.

\begin{table}[h]
\begin{tabular}{|c|c|c|c|c|c|}\hline
$K$ & $t$-range & $m_P$ & $m_P^{\prime}$ & $f_P/Z_A$
& $\chi^2/{\rm dof}$ \\
\hline
0.14144 & 7-23  & 0.297(1.7) & 0.679(41) & 0.0620(14) & 195/194 \\
 mixed & 7-23 & 0.258(2.6)   & 0.628(63) & 0.0566(15) & 183/194 \\
0.14226 & 7-23 & 0.212(5)   & 0.530(104) & 0.0508(24) & 181/194 \\
``K'' & &0.182(6) & &0.0476(20) & \\
\hline
$K$ & $t$-range & $m_V$ & $m_V^{\prime}$ & $1/(f_V Z_V)$
 & $\chi^2/{\rm dof}$ \\
\hline
0.14144 & 7-23  & 0.389(4) & 0.835(45) & 0.311(6) & 42/45 \\
mixed & 7-23  & 0.365(5) & 0.815(49) & 0.322(7) & 34/45 \\
0.14226 & 7-23  & 0.340(7) & 0.796(55) & 0.336(9) & 27/45 \\
``K$^*$'' & &0.327(11) & &0.343(9) & \\
\hline
$K$ & $t$-range & $m_A$ & $m_A^{\prime}$ & $1/(f_A Z_A)$
 & $\chi^2/{\rm dof}$ \\
\hline
0.14144 & 4-15  & 0.569(10) & 1.0 & 0.178(19) & 15/19 \\
0.14226 & 4-15  & 0.534(15) & 1.0 & 0.195(20) & 15/19 \\
``K$_A^*$'' & &0.524(17) & &0.200(21) & \\
\hline
\end{tabular}
 \caption{ The pseudoscalar, vector and axial vector meson masses and
couplings  in lattice units (``mixed'' refers to mesons with one quark
of each hopping parameter $K$). The rows marked K, K$^*$ and
K$_A^*$ refer  to an extrapolation to a quark mass such that $m_V=1.8
m_P$. }
 \end{table}

 From these values we can calculate the dimensionless combination
$J=0.37(3)$ where the error here is statistical. The main source  of
error in the determination of $J$ comes from the difference  of vector
meson masses ($am_V(q_1)-am_V(q_2)$).  The systematic error  on this
vector meson mass difference of 0.037 from using different $t$-ranges
and different  correlated fitting procedures is $\pm 0.002 $ which is 5
\%. Thus  one can express the error on $J$ by adding a second systematic
error as $J=0.37(3)(2)$.  As can be seen from fig.~2, the UKQCD data  on
$m_V$ versus $m_P^2$ is consistent with a straight line but  also allows
some small curvature. This introduces a further  systematic error since
the slope and intercept are required at  $m_V/m_P=1.8$ which is an
extrapolation from the three smeared-source  data points used above.
This source of systematic error is hard to quantify since statistically
significant evidence for any curvature does not exist. We suggest that
$J=0.37(3)(4)$ is an appropriate final estimate from our quenched
lattice study.

This value is significantly different from the experimental value
of $J=0.48(2)$ given above.   Since $J$ does not depend on the lattice
spacing value used, nor on the critical hopping parameter $K_{crit}$,
the discrepancy points towards an underlying difference between
the quenched lattice and the full QCD experimental values.

 There are also published data for the clover fermionic action at
$\beta=6.2$ from the APE collaboration~\cite{APE62} using a $18^3 \times
64$ lattice. Their results for $am_V$ and $am_P$ versus $K^{-1}$
disagree with those published by UKQCD~\cite{ukqcd} and  confirmed
above. The APE results were obtained from  a one state fit (plateau fit)
to the local correlators (LL in our  notation). Such a fit can be
influenced by inadequate  statistics since the signal is relatively
noisy at the  large $t$-values (15 to 28) where the plateau in the
effective mass  is determined -  see, for example the discussion in
ref~\cite{cmlat94}.

The disagreement in meson mass values is more  pronounced at lower quark
masses where the APE results for $am$ are  lower than the interpolated
UKQCD values - see fig.~2. This  might be caused by enhanced finite
size effects (at spatial size $18^3$ compared to $24^3$) at such light
masses.  Such a conclusion is, however, in conflict with the  results
for staggered fermions~\cite{KT} and Wilson fermions~\cite{GF11} that
the $am$ values are increased on a smaller spatial volume. A resolution
of this disagreement would need higher statistics and/or the
measurement of further correlations (such as smeared correlations) at
the smaller volume.

Returning now to the axial vector meson spectrum, our results give $
m_A/m_V=1.60(7)$ at a quark mass corresponding to $m_V=1.8m_P$. This
can be compared with the value of the ratio deduced from experimental
data: 1.48(2) as discussed above. Again there is evidence for a
discrepancy between the quenched lattice result and experiment. Since
there has been relatively little study of this ratio in lattice work,
the finite size and finite $a$ errors are not easy to estimate. Further
work is again needed.

\section{Discussion}

Even though $J$ does not depend on the lattice spacing explicitly (it
can be thought of as a mass ratio), it will  have
discretisation errors. These will be of order $a$ for the Wilson
fermionic action and of order $\alpha_S a$ for the clover fermionic
action. The other lattice approximation that needs to be checked is
the finite lattice size used.

Within the UKQCD collaboration we intend to study the discretisation
and finite-size errors systematically. At present, we have data at  one
lattice size ($24^3 \times 48$) and at one $\beta$-value (6.2)  only.
We chose these lattice specifications with the expectation that  the
discretisation and finite size errors would be very small for  the light
meson spectrum.  To explore further the possible discretisation and
finite size errors, we make some comparisons with other lattice
determinations  of the meson spectrum. To make a comparison  using the
clover fermionic action, we analyse the APE data~\cite{APE}  at
$\beta=6.0$. Making a fit to their published results for $m_V$ versus
$m_P^2$ for three light $K$-values, we obtain $J=0.37(2)$. We do not
have access to the  correlations among their results at different
$K$-values, so the  error estimate is based on an uncorrelated fit with
the quoted diagonal  errors only. Their lattice spatial size ($18^3$) is
comparable in  physical volume to our work at $\beta=6.2$. Thus we can
conclude that we  see no sign of any $a$ dependence in $J$ as determined
from quenched clover  lattices. This is confirmation of our assumption
above that the clover action  should remove most of the $a$ dependence
for $\beta > 6.0$.

There are more extensive lattice data available for Wilson fermions. In
this case, one would expect somewhat larger $a$ dependence to  remain
than in the clover case just discussed. We consider the extraction  of
$J$ from this data.  One of the most comprehensive studies of the light
meson spectrum  has been made by the GF11 group~\cite{GF11}. Their
largest $\beta$-value is  6.17 with lattice size
$32^2\times30\times40$.  From their published results for  $ (m_p^2
/m_v)\, d m_V / d m_P^2$, one can evaluate $J=0.373(13)$. This result is
similar to the clover  result for the nearby $\beta$-value of 6.2. The
GF11 group has also studied a sequence of lattices with  coarser lattice
spacing but similar physical size.  The results for $J$ are shown in
fig~8 by squares. The errors  are sufficiently large that the results
are compatible either with essentially no discretisation  error ($J$
independent of $a$) or with a substantial linear $a$  dependence which
allows the experimental value to be obtained as the  $a \to 0$ limit.

The GF11 data at $\beta=5.7$ have two different  lattice spatial
volumes. This enables the finite size effects to be explored. Again the
data  are consistent with no  volume  dependence of $J$ - but have
sufficiently large errors that they  also would allow quite large
volume dependence in principle. Using data at as low a $\beta$ value as
5.7  to determine the order $a$ finite size effects from Wilson fermions
is also  rather uncontrolled since the order $a^2$ discretisation
effects from  the pure gauge sector are known to be significant (over
30\% in glueball to string tension ratio~\cite{gb}) at 5.7.

 A study of finite-size effects can also be made by comparing the data
at $\beta=6.0$ with Wilson fermions from the LANL group~\cite{LANL},
QCDPAX group~\cite{QCDPAX} and APE group~\cite{APE}.   The plot of $m_V$
against $m_P^2$ from these data is shown in fig.~1. The spectra with
different spatial lattice volumes ($32^3, 24^3$ and $18^3$ respectively)
are seen to be  in  agreement which implies  that $J$ is independent of
lattice spatial volume.  The most comprehensive study is by the LANL
group and  fitting their data for $m_V$ versus  $m_P^2$ yields
$J=0.38(1)$. This result also confirms that there is no  substantial
finite size effect in the determination of $J$ since it agrees well with
the GF11 data from smaller physical volumes at  $\beta=6.17$ and 5.93.

 Since the clover fermionic action is expected to have smaller
discretisation errors than the Wilson action, the fact that the clover
and Wilson discretisation results for $J$  agree for $\beta \ge 6.0$,
suggests that there can be no substantial  $a$ dependence in either.
Furthermore all results, except the smaller volume GF11 determination  at
$\beta=5.7$, agree with $J=0.37$.   Further precision data are needed to
reinforce this conclusion that $J=0.37$ for quenched QCD.

To measure $J$ directly in lattice simulation with dynamical fermions
will  be hard. The usual procedure for dynamical fermion studies is to
use sea quarks of one mass  in such work. Then only $am_V(q_1,q_1)$ and
$am_P(q_1,q_1)$ would  be available. Making a direct comparison of
meson masses at different dynamical quark mass values is  inappropriate
since the lattice spacing $a$ will depend on the sea-quark mass. A way
forward is to use a partly quenched approach  with a fixed sea-quark
mass but several valence quark masses.  Typical results~\cite{df} use
two flavours of staggered fermions  as sea-quarks and Wilson valence
quarks. The analysis of those data  ($16^3 \times 32 $ lattice at
$\beta=5.6$ with $am_q=0.01$ and 0.025) yields $J=0.356(11)$ and
$J=0.345(7)$ respectively from fitting the published  mass values for
$m_V$ versus $m_P^2$. These values are  similar to those obtained above
from quenched lattices.  Indeed the authors point out that their results
for light hadrons are indistinguishable from those in the quenched
approximation. Thus it appears that the full QCD result will only be
approached when the sea-quark masses used in the lattice studies are
reduced further.

To explore  dynamical quark simulation fully will need much more  work.
Implementing two different sea-quark masses will be complicated.
Using realistic masses will be even harder. There  are also
indications  of enhanced finite-size  effects in dynamical quark
studies.

\section{Conclusions}

Quenched QCD is an excellent model for the hadronic interactions.  It is
an approximation and the accuracy of the approximation can  be estimated
in some cases. For instance, the beta-function can be  evaluated
perturbatively and the quenched approximation amounts  to treating
$(33-2N_f)$ as 33, an error of 20\%. The width of the  $\rho$ meson is
zero in the quenched approximation, in contrast  to the non-zero value
in full QCD.  Previous lattice studies  in quenched QCD have claimed
agreement~\cite{GF11} with the  experimental light hadron spectrum. In
this work, we have  concentrated on a quantity $J$ which can be
determined from the  pseudoscalar and vector meson masses and which is
independent  of the lattice spacing and hopping parameters.  We find
evidence that $J=0.37(2)(4)$ from quenched lattice studies. This is  in
contrast to the experimental value of $J=0.48(2)$. Further  lattice
evaluations are needed to confirm the assumptions we  have made about
the discretisation and finite size errors on $J$. We also have presented
some  evidence for a discrepancy between the  ratio of axial and vector
meson masses from quenched lattices compared  to experiment.

These discrepancies should be no cause for surprise. What it does
show is that a careful study of lattice QCD with dynamical
quarks is of importance if an accurate description is needed.

\section{Acknowledgements}

This research was supported by the UK Science and Engineering Research
Council under grants GR/G~32779, GR/H~49191, GR/J~21200 and GR/H~53624,
by the University of Edinburgh and by Meiko Limited.  We are grateful to
Edinburgh University Computing Service and, in particular, to Mike Brown
for maintaining service on the Meiko i860 Computing Surface. We
thank our colleagues in the UKQCD collaboration at Edinburgh for
their assistance in making the data sets available to us. We
acknowledge helpful advice from Chris Sachrajda and Hartmut Wittig.

\newpage

\begin{figure}[h]
\vspace{16cm}
\includegraphics{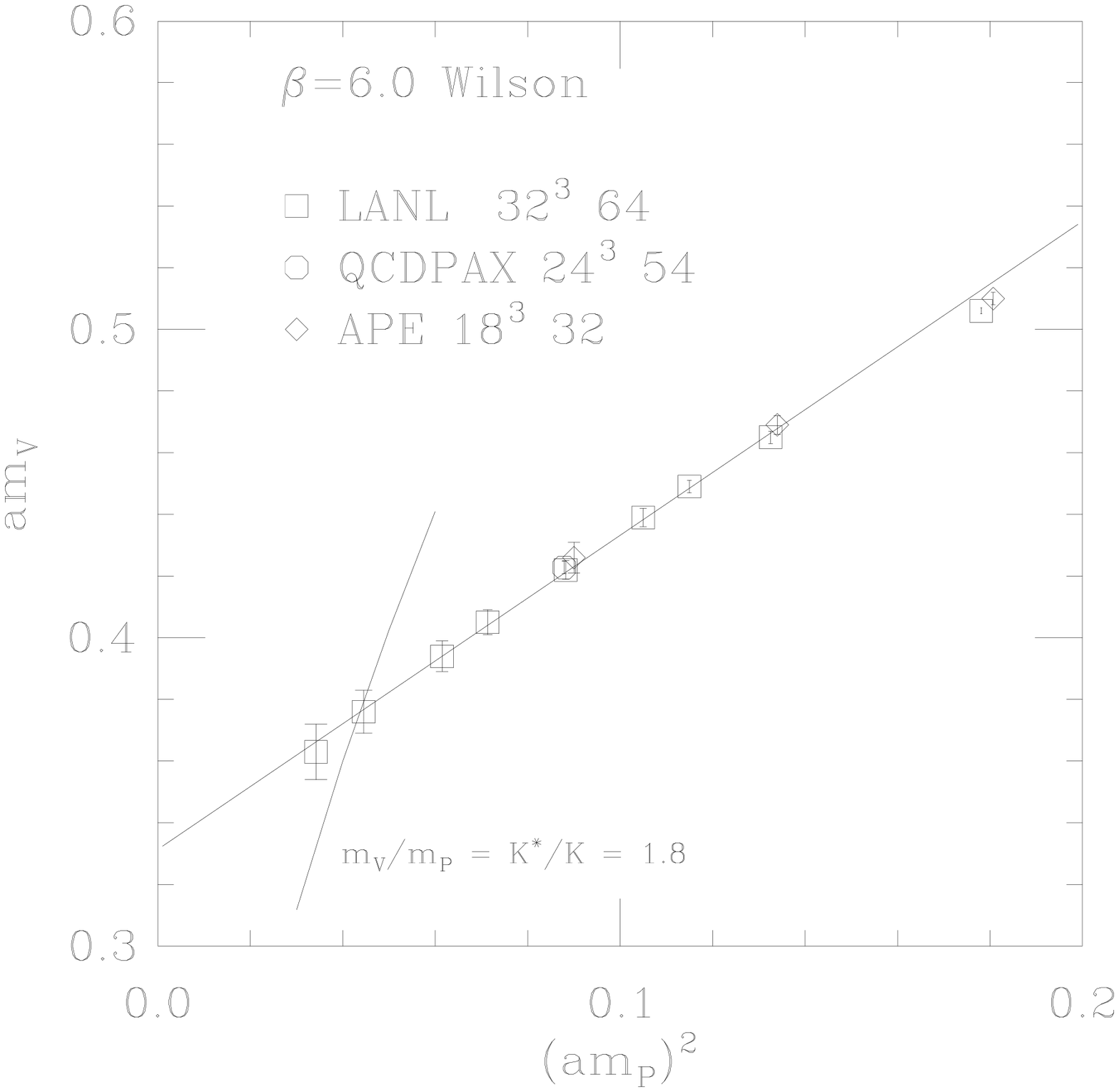}
 \caption{ The value of $am_V$ versus $(am_P)^2$
in lattice units.  The data are for Wilson fermions at $\beta=6.0$ from
the LANL, QCDPAX and APE groups. The LANL data have pairs of unequal
quark masses as well as equal.  A straight line fit to the LANL data,
excluding the heaviest mass point, is shown. The curve corresponds
to the reference quark mass scale set by $m_V=1.8 m_P$.
 }
\end{figure}

\begin{figure}[h]
\vspace{16cm}
\includegraphics{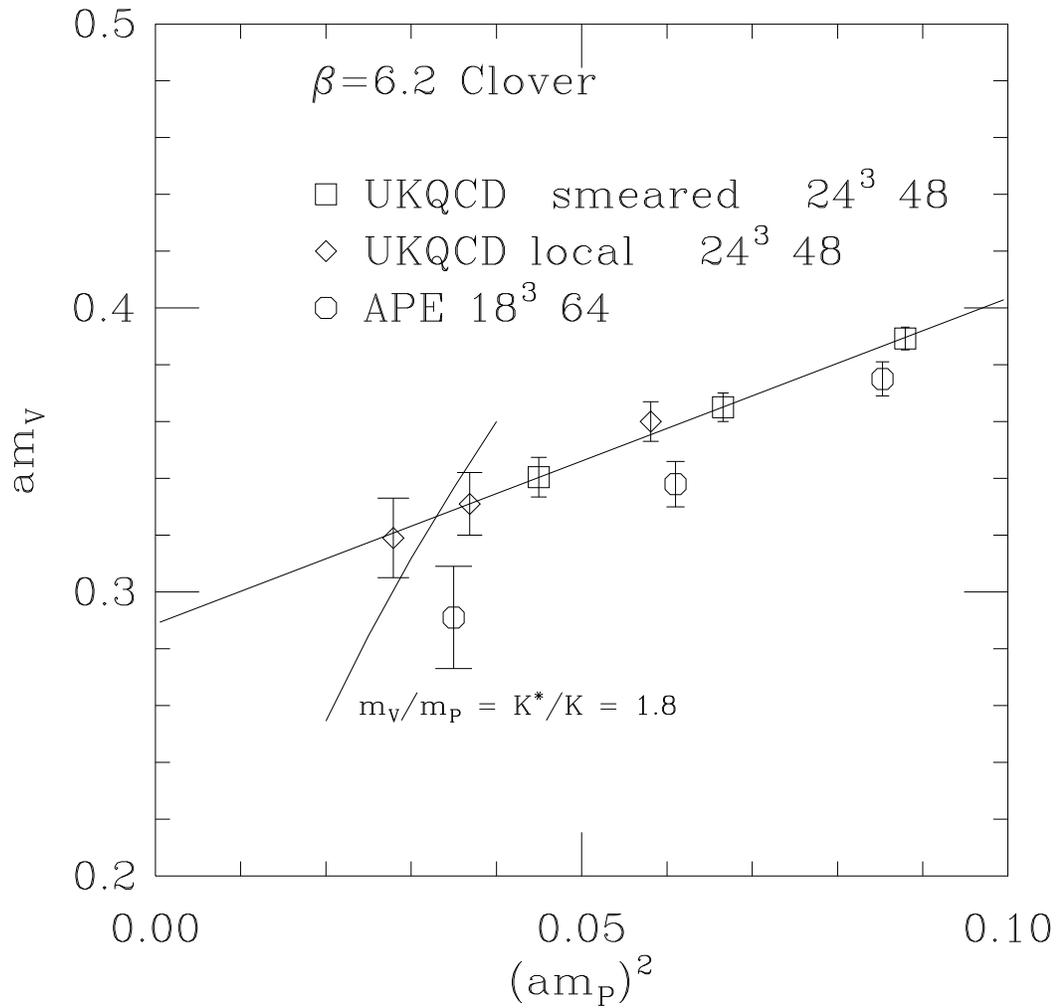}
 \caption{ The value of $am_V$ versus $(am_P)^2$   in lattice units.
The data are for the clover fermionic action at $\beta=6.2$ from  the
UKQCD and APE groups. The UKQCD data have pairs of unequal  quark masses
as well as equal.  A straight line fit to the UKQCD data from  smeared
sources (squares)  is shown. The curve corresponds  to the reference
quark mass scale set by $m_V=1.8 m_P$.
 }
\end{figure}

\begin{figure}[h]
\vspace{16cm}
\includegraphics{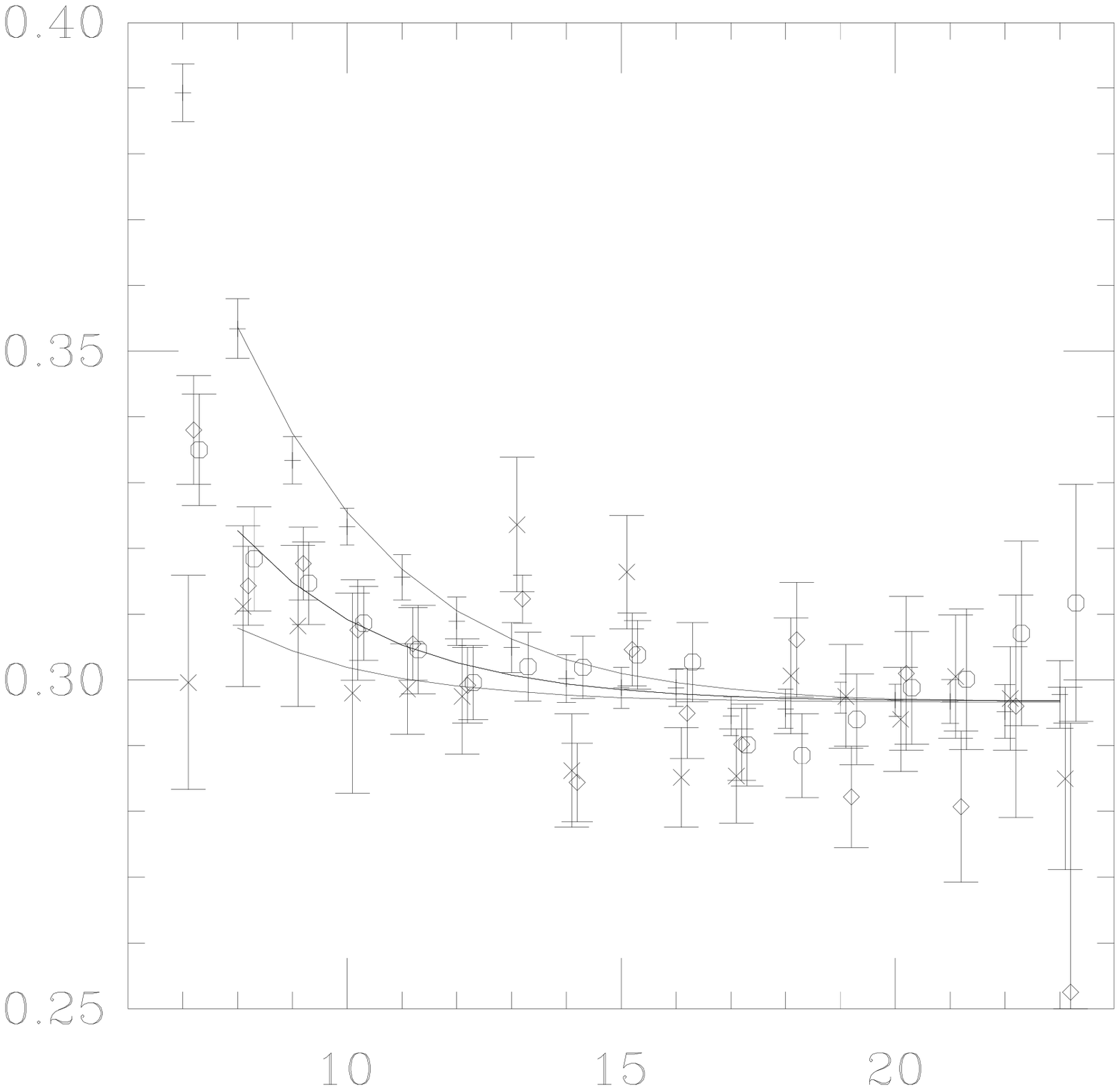}
\caption{ The effective pseudoscalar mass in lattice units from
local sinks and sources (LL) versus $t$ for hopping parameter
$K=0.14144$. The symbols are for spin
combinations at source and sink PP (+), AA ($\times$), AP(octagon) and
PA(diamond). The curves represent the two-state factorising fits
described in the text.
 }
\end{figure}

\begin{figure}[h]
\vspace{16cm}
\includegraphics{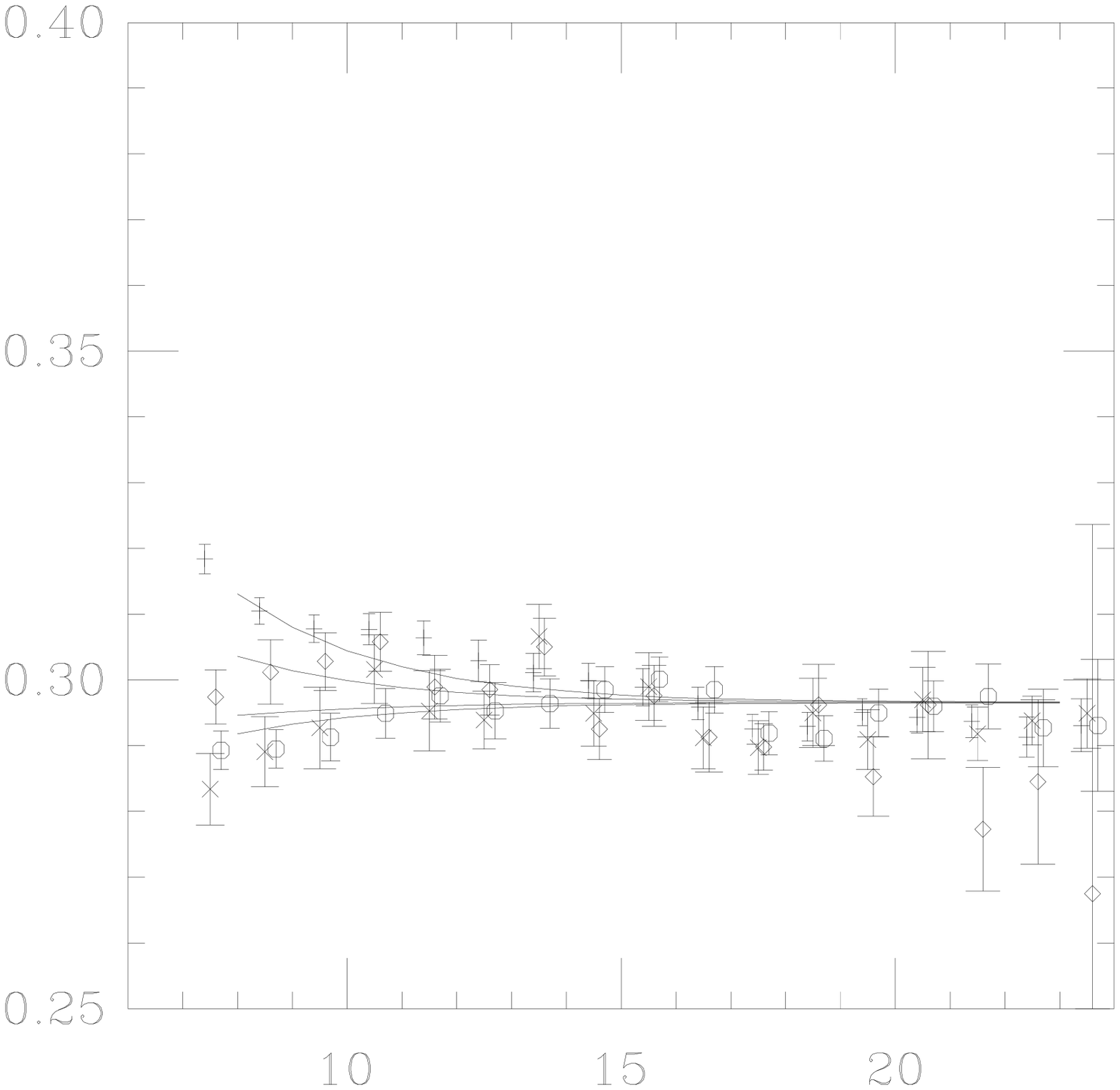}
\caption{ The effective pseudoscalar mass in lattice units from
local sinks and smeared source (SL) versus $t$ for hopping parameter
$K=0.14144$. The symbols are for spin
combinations at source and sink PP (+), AA ($\times$), AP(octagon) and
PA(diamond). The curves represent the two-state factorising fits
described in the text.
 }
\end{figure}

\begin{figure}[h]
\vspace{16cm}
\includegraphics{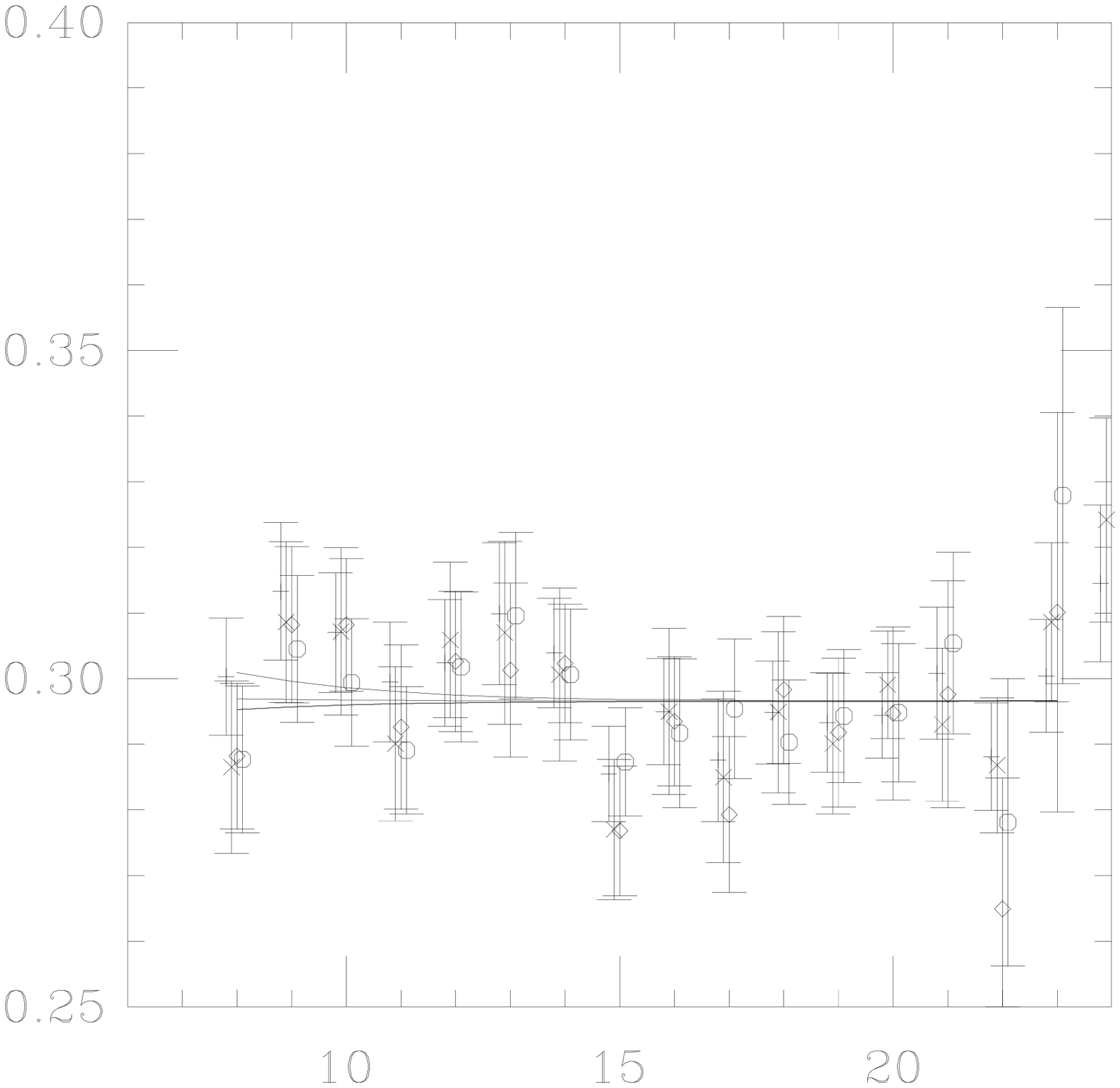}
\caption{ The effective pseudoscalar mass in lattice units from
smeared sinks and sources (SS) versus $t$ for hopping parameter
$K=0.14144$. The symbols are for spin
combinations at source and sink PP (+), AA ($\times$), AP(octagon) and
PA(diamond). The curves represent the two-state factorising fits
described in the text.
 }
\end{figure}

\begin{figure}[h]
\vspace{16cm}
\includegraphics{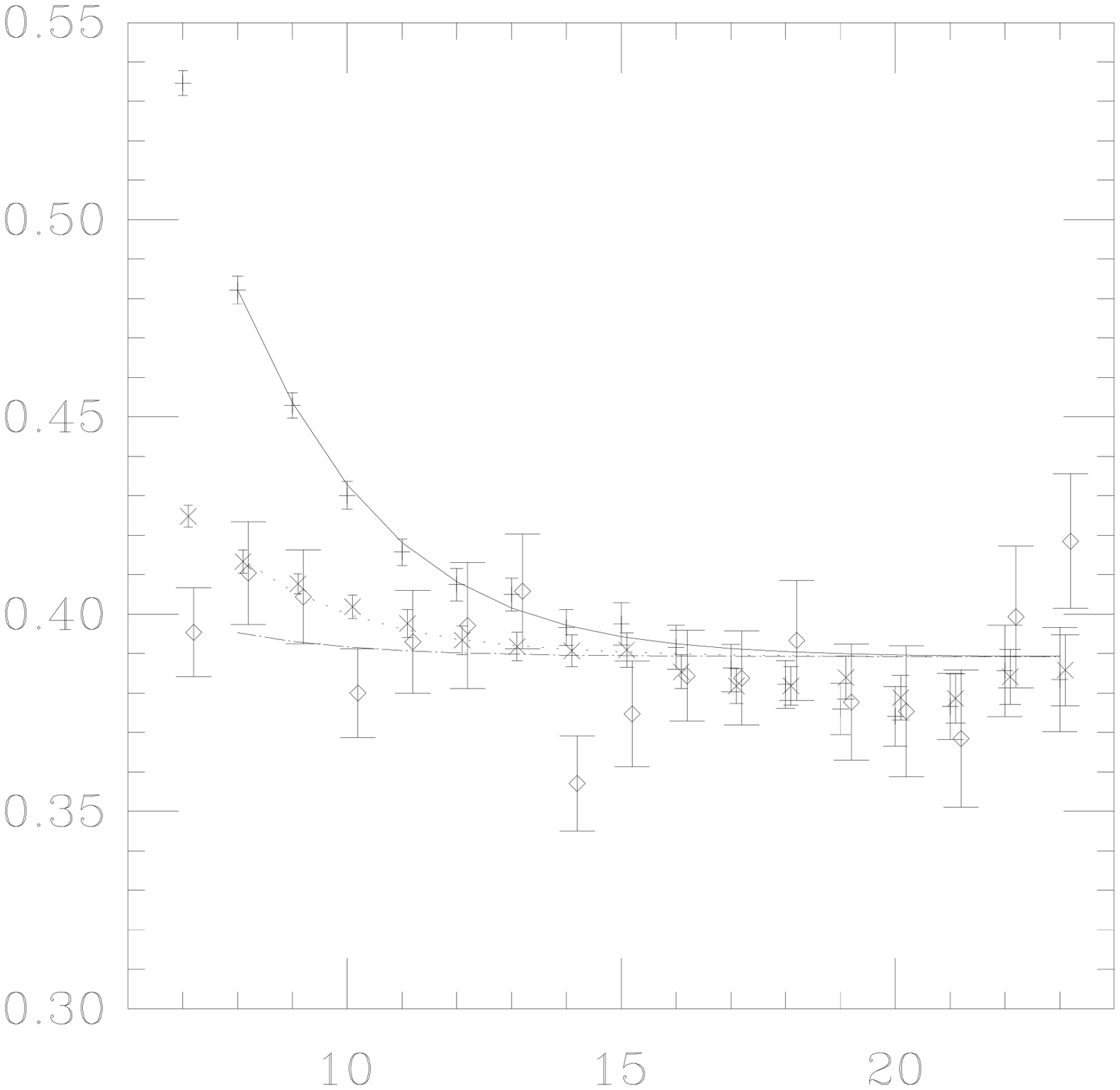}
\caption{ The effective vector mass in lattice units from
local and smeared sinks and sources (SS) versus $t$ for hopping parameter
$K=0.14144$. The symbols are for operators at source and sink
LL (+), SL ($\times$), SS(diamond). The curves represent the two-state
factorising fits  described in the text.
 }
\end{figure}

\begin{figure}[h]
\vspace{16cm}
\includegraphics{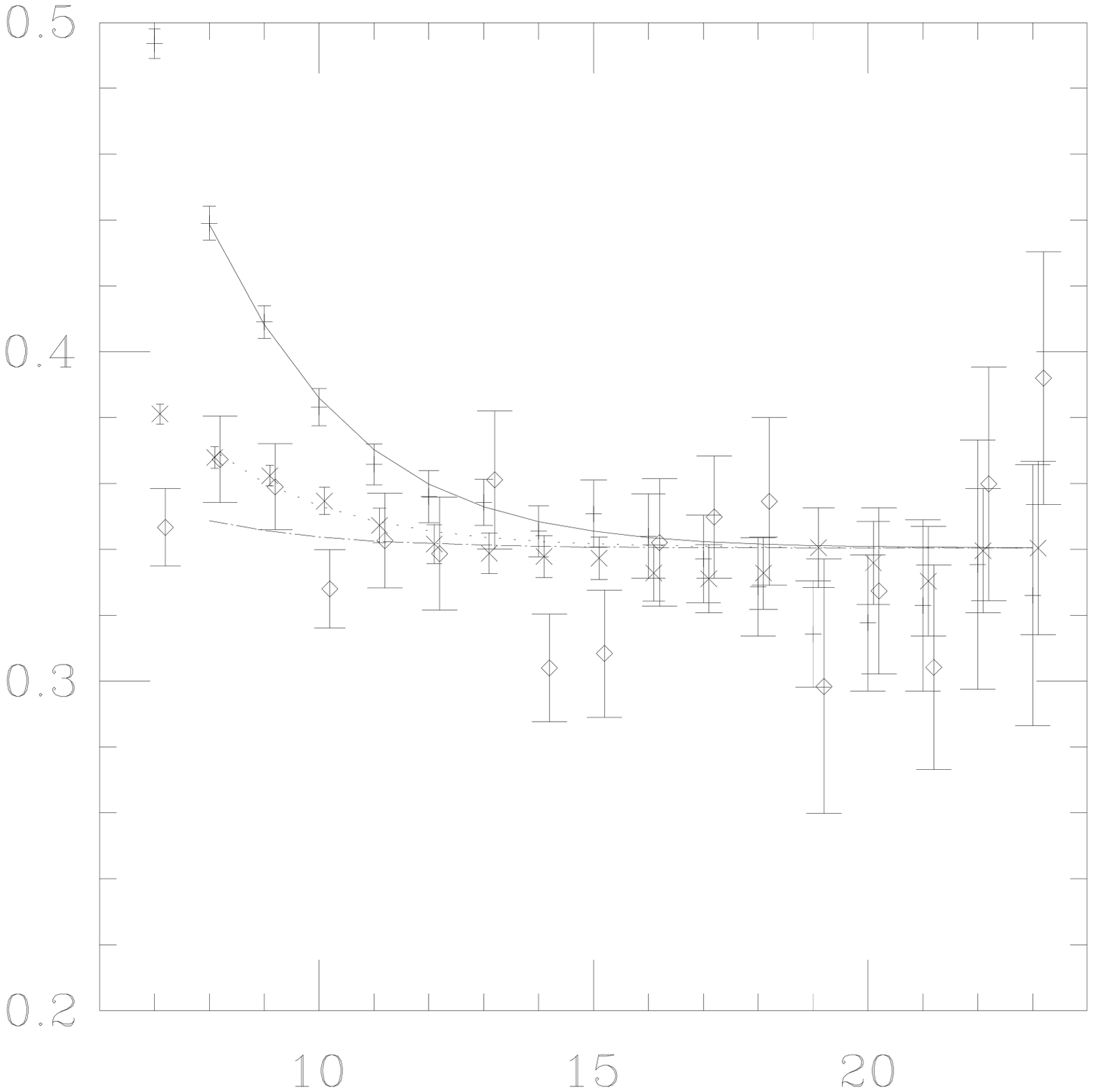}
\caption{ The effective vector mass in lattice units from
local and smeared sinks and sources (SS) versus $t$ for hopping parameter
$K=0.14226$. The symbols are for operators at source and sink
LL (+), SL ($\times$), SS(diamond). The curves represent the two-state
factorising fits  described in the text.
 }
\end{figure}

\begin{figure}[h]
\vspace{16cm}
\includegraphics{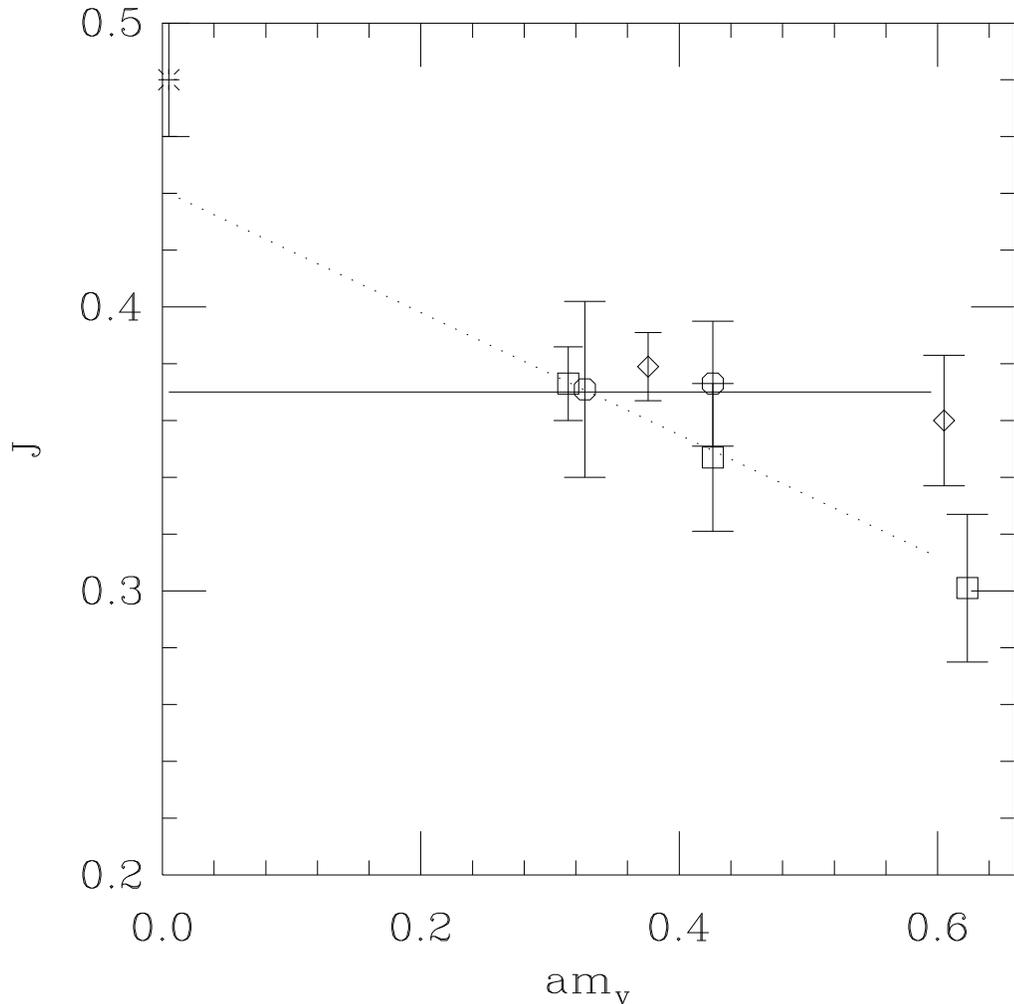}
 \caption{ The value of $J=m_v dm_V/dm_P^2$
 versus $m_v$ in lattice units.  The data from clover fermions are shown
by  octagons from the present work (with statistical error only) and
from the APE group. The data from Wilson fermions are shown  by
squares  from the GF11 group  at similar physical spatial volumes and
diamonds for the  larger volume results from GF11 and LANL. The
cross is the experimental  value. %ref~\protect\cite{gf11}.
 The solid  and dotted lines represent different interpretations of the
data,  as discussed in the text.
 }
\end{figure}

\end{document}